\documentclass[letterpaper]{jpconf}

\usepackage{graphicx}
\usepackage{epsfig}

\begin{document}

\setlength{\topmargin}{-1cm}

\title{A Metric for Testing the Nature of Black Holes}

\author{ Tim Johannsen\footnote{CITA National Fellow} }

\address{ Department of Physics and Astronomy, University of Waterloo, 200 University Avenue West, \\
Waterloo, ON, N2L 3G1, Canada \\
Perimeter Institute for Theoretical Physics, 31 Caroline Street North, \\
Waterloo, ON, N2L 2Y5, Canada \\
Physics Department, University of Arizona, 1118 East 4th Street, Tucson, AZ 85721, USA
}

\begin{abstract}

In general relativity, astrophysical black holes are uniquely described by the Kerr metric. Observational tests of the Kerr nature of these compact objects and, hence, of general relativity, require a metric that encompasses a broader class of black holes as possible alternatives to the usual Kerr black holes. Several such Kerr-like metrics have been constructed to date, which depend on a set of free parameters and which reduce smoothly to the Kerr metric if all deviations vanish. Many of these metrics, however, are valid only for small values of the spin or small perturbations of the Kerr metric or contain regions of space where they are unphysical hampering their ability to properly model the accretions flows of black holes. In this paper, I describe a Kerr-like black hole metric that is regular everywhere outside of the event horizon for black holes with arbitrary spins even for large deviations from the Kerr metric. This metric, therefore, provides an ideal framework for tests of the nature of black holes with observations of the emission from their accretion flows, and I give several examples of such tests across the electromagnetic spectrum with current and near-future instruments.

\end{abstract}

\section{Introduction}

By virtue of the no-hair theorem, black holes in general relativity are fully and uniquely characterized by only three parameters: their masses, spins, and electrical charges (Israel 1967, 1968; Carter 1971, 1973; Hawking 1972; Robinson 1975; Mazur 1982). These objects are described by the Kerr-Newman metric (Newman et al. 1965), which, in the case of astrophysical black holes, reduces to the Kerr metric (Kerr 1963), since any net charge quickly neutralizes.

The no-hair theorem, then, provides a means to test general relativity by verifying that black holes are indeed Kerr black holes. Several such tests have been suggested to date, which are based on observations of either gravitational waves (see Hughes 2010 for a review), electromagnetic emission from accretion flows (Johannsen \& Psaltis 2010a), pulsar-black hole binaries (Wex \& Kopeikin 1999), the ephemeredes of stars around the supermassive black hole Sgr~A* at the galactic center (Will 2008), or of the quasar OJ287 (Valtonen et al. 2011).

While tests in the regime of weak gravitational fields can rely on a parameterized post-Newtonian formulation (e.g., Will 1993), strong-field tests require a careful modeling of the black hole spacetime itself in terms of a parametric deviation from the Kerr metric. These Kerr-like spacetimes depend on one or more free parameters in addition to the mass and spin, and signatures of possible violations of the no-hair theorem can be studied as a function of the deviation parameters. Several parametric deviations have been constructed so far (e.g., Manko \& Novikov 1992; Collins \& Hughes 2004; Glampedakis \& Babak 2006; Yunes \& Pretorius 2009; Vigeland \& Hughes 2010; Vigeland, Yunes, \& Stein 2011; Johannsen \& Psaltis 2011b; Yagi, Yunes \& Tanaka 2012).

In this paper, I review the Kerr-like black hole spacetime constructed by Johannsen \& Psaltis (2011b), which is regular everywhere outside of the event horizon even for high values of the spin and large deviations from the Kerr metric. This property makes this metric an ideal framework for strong-field tests the no-hair theorem with observations of the accretion flows of black holes.

\section{A Suitable Metric for Tests of the No-Hair Theorem with Observations of Black Hole Accretion Flows}

The Kerr metric is a stationary, axisymmetric, asymptotically flat, vacuum solution of the Einstein field equations and, due to the no-hair theorem, the only spacetime in general relativity that has all of these characteristics. Kerr-like metrics retain as many of these properties in order to closely mimic the observational appearance of Kerr black holes. All of them include the Kerr metric as the limiting case if the deviation parameters are set to zero.

By construction, parametric deviations from the Kerr metric are usually stationary, axisymmetric, and asymptotically flat. If such a metric is also a vacuum solution in general relativity, it either harbors a naked singularity or is plagued with pathological regions in the exterior domain where causality is violated (see Johannsen et al. 2012a). Some Kerr-like metrics are valid only for small or intermediate values of the black hole spin (e.g., Glampedakis \& Babak 2006; Yunes \& Pretorius 2009), while others are based on an expansion in the deviation parameters in order to remain a vacuum solution of the Einstein equations under certain conditions (e.g., Vigeland \& Hughes 2010). Depending on the desired application, additional metric properties can be important, such as its Petrov type (Vigeland et al. 2011) or violations of parity (Yunes \& Pretorius 2009; Yagi et al. 2012).

The metric by Johannsen \& Psaltis (2011b) in Boyer-Lindquist-like coordinates is given by the line element
\begin{eqnarray}
ds^2 &=& -[1+h(r,\theta)] \left(1-\frac{2Mr}{\Sigma}\right)dt^2 -\frac{ 4aMr\sin^2\theta }{ \Sigma }[1+h(r,\theta)]dtd\phi + \frac{ \Sigma[1+h(r,\theta)] }{ \Delta + a^2\sin^2\theta h(r,\theta) }dr^2 \nonumber \\
&& + \Sigma d\theta^2 + \left[ \sin^2\theta \left( r^2 + a^2 + \frac{ 2a^2 Mr\sin^2\theta }{\Sigma} \right) + h(r,\theta) \frac{a^2(\Sigma + 2Mr)\sin^4\theta }{\Sigma} \right] d\phi^2,
\label{metric}
\end{eqnarray}
where
\begin{equation}
\Sigma \equiv r^2 + a^2 \cos^2\theta,~~~\Delta \equiv r^2 - 2Mr + a^2,~~~h(r,\theta) \equiv \sum_{k=0}^\infty \left( \epsilon_{2k} + \epsilon_{2k+1}\frac{Mr}{\Sigma} \right) \left( \frac{M^2}{\Sigma} \right)^{k}.
\end{equation}

This metric is a stationary, axisymmetric, and asymptotically flat parametric deviation that describes an actual black hole for all values within the range of spins $|a| \leq M$ without relying on an expansion in the deviation parameters $\epsilon_k$ and that is regular, i.e., it is free of any pathologies outside of the event horizon. Possible nonzero lowest-order deviations $\epsilon_k$, $k=0,1,2$, of the Kerr metric are ruled out by current solar-system tests of gravity and by the requirement that the above metric reduces to Newtonian gravity far from the black hole. For simplicity, I will allow the leading-order coefficient $\epsilon_3$ to be the only nonvanishing deviation parameter.

\begin{figure}[h]
\begin{center}
\psfig{figure=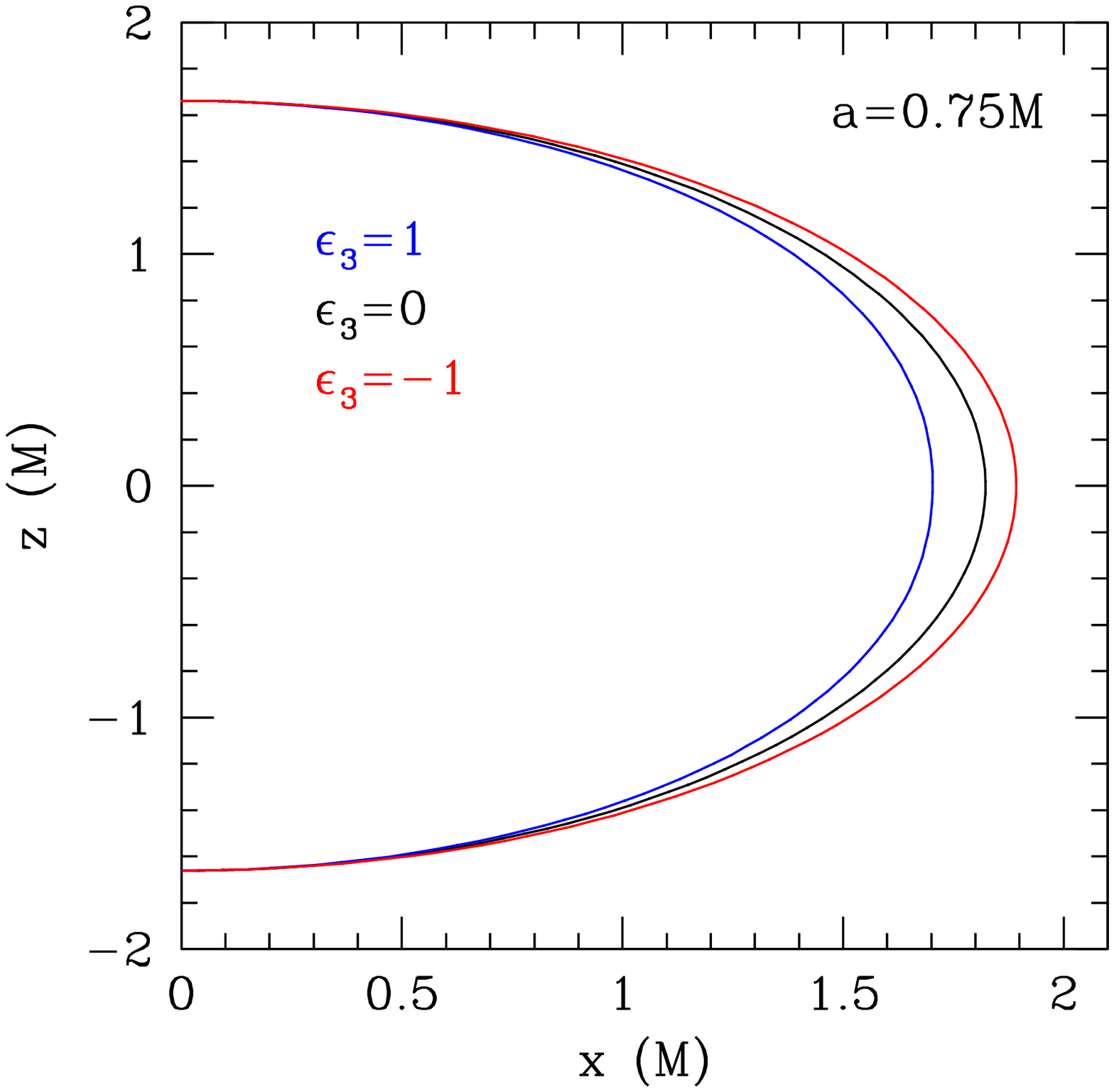,height=3.1in}
\psfig{figure=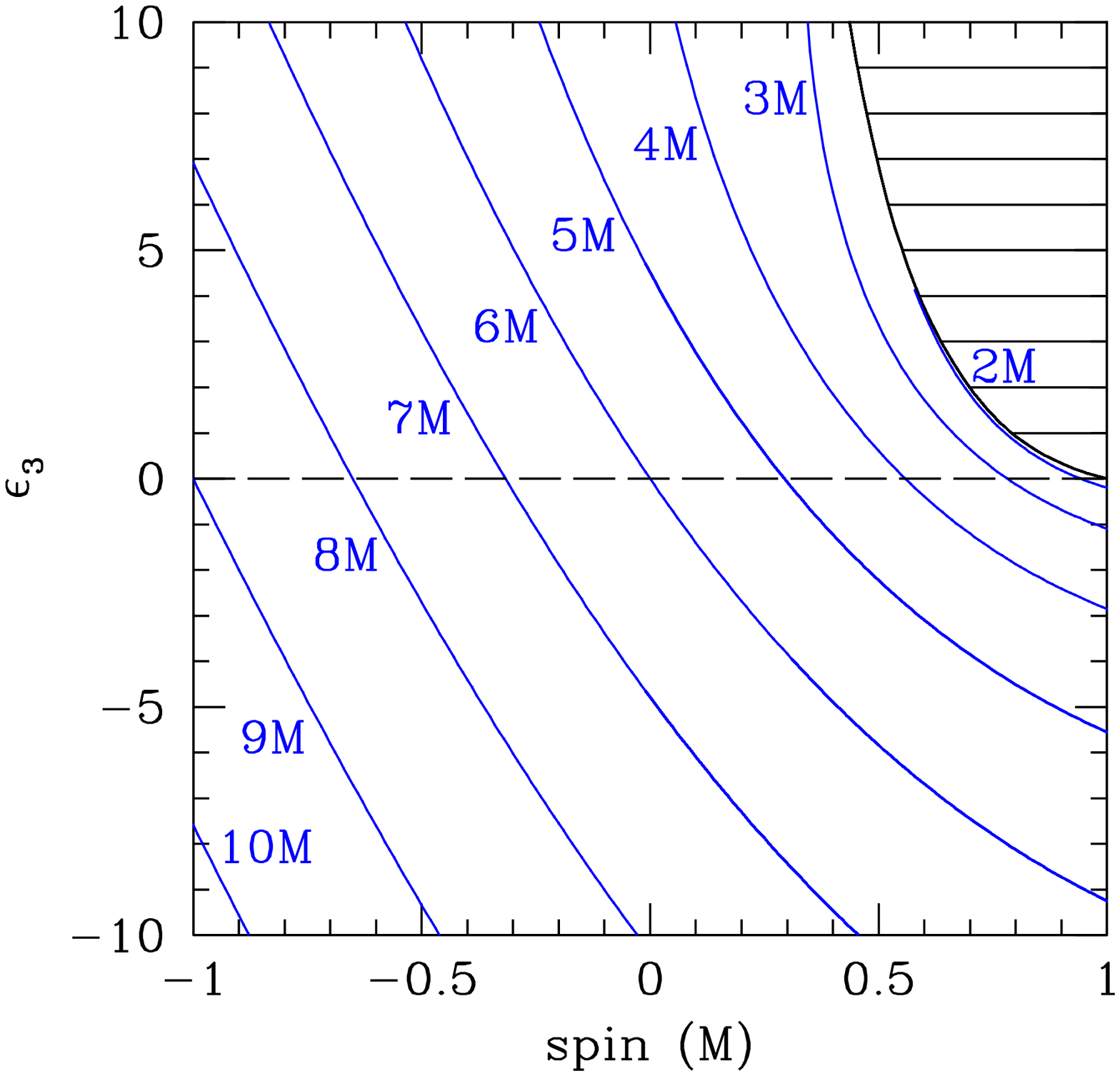,height=3.1in}
\end{center}
\caption{Left: Event horizon of a black hole with spin $a=0.75M$ for several values of the parameter $\epsilon_3$ that measures the deviation from the Kerr metric. For values of the parameter $\epsilon_3>0$ the shape of the event horizon is prolate, while for values of the parameter $\epsilon_3<0$ the shape of the event horizon is oblate. Right: Contours of constant radius of the ISCO for values of the spin $-1 \leq a/M \leq 1$ and of the deviation parameter $-10 \leq \epsilon_3 \leq 10$. The ISCO radius decreases for increasing values of the spin and the parameter $\epsilon_3$. The shaded region marks the part of the parameter space where the event horizon is not closed (only shown for the case $a>0$; Johannsen \& Psaltis 2011b).}
\label{properties}
\end{figure}

With this choice, the shape of the event horizon depends on the values of the spin and the parameter $\epsilon_3$. Figure~\ref{properties} shows the location of the horizon in the $xz$-plane of a black hole with a value of the spin $a=0.75M$, where $x \equiv \sqrt{r^2 + a^2} \sin \theta$ and $z \equiv r \cos \theta$. While the event horizon of a Kerr black hole is spherical, the horizon of a black hole in this metric is prolate for positive values of the parameter $\epsilon_3$ and oblate for negative values of the parameter $\epsilon_3$. For high values of the spin and values of the parameter $\epsilon_3$ exceeding a maximum bound as shown in Figure~\ref{properties}, the central object turns into a naked singularity.

The properties of this spacetime make it ideally suited for tests of the no-hair theorem that involve observations of accretion flows. The innermost stable circular orbit (ISCO) is often taken to be the inner edge of geometrically thin accretion disks and can be arbitrarily close to the event horizon in the case of rapidly spinning black holes. Since relativistic effects are most prominent in the immediate vicinity of black holes, it is critical that this region is properly modeled by a parametric deviation. Since the metric by Johannsen \& Psaltis (2011b) is regular everywhere outside of the horizon, the electromagnetic emission of accretion flows can be modeled self-consistently. Figure~\ref{properties} also shows the location of the ISCO as a function of the spin and the parameter $\epsilon_3$. For increasing values of the spin and the deviation parameter, the radius of the ISCO approaches the event horizon.

\section{Astrophysical Tests}

\begin{figure}[ht]
\begin{center}
\psfig{figure=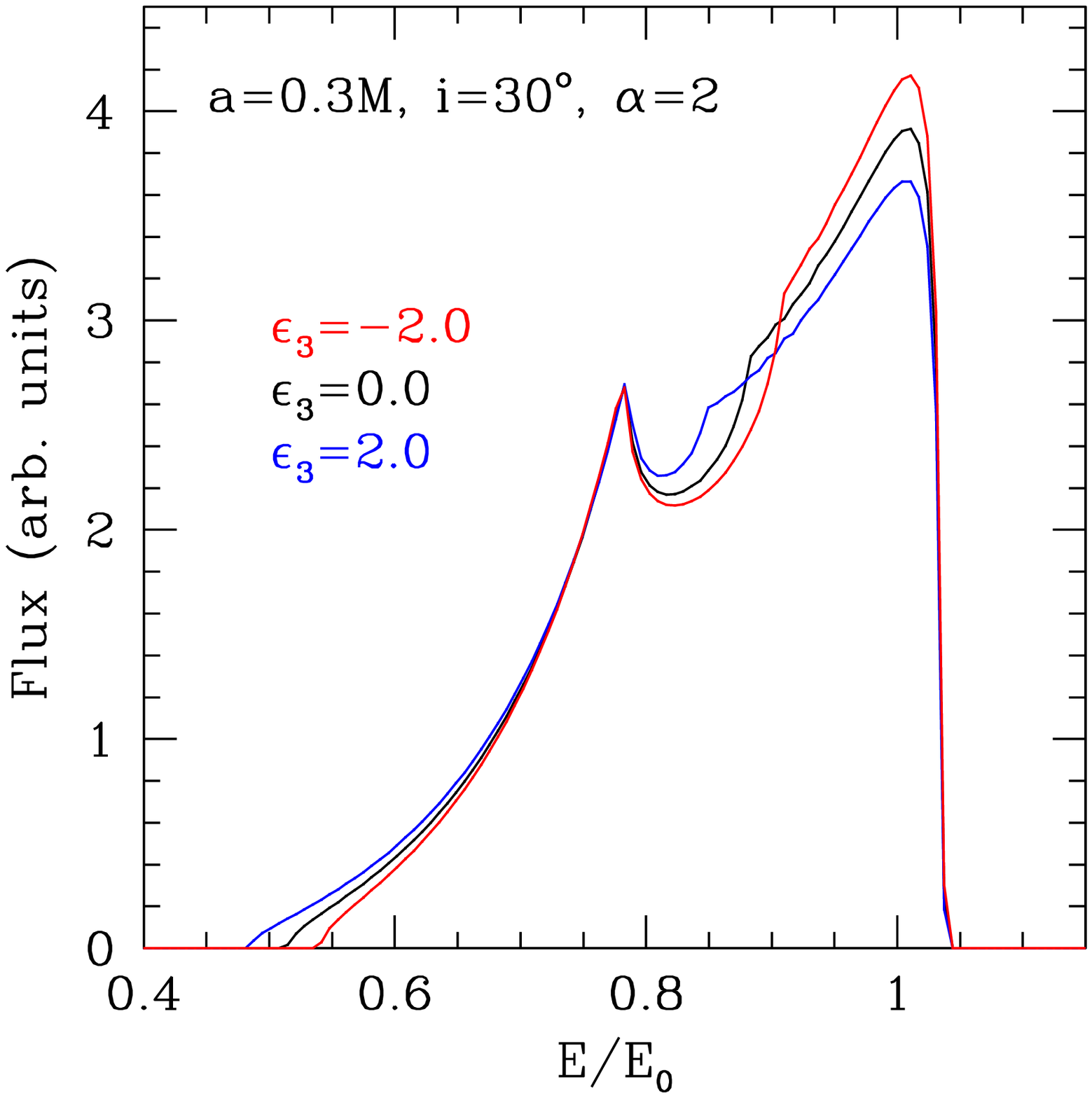,height=3.1in}
\psfig{figure=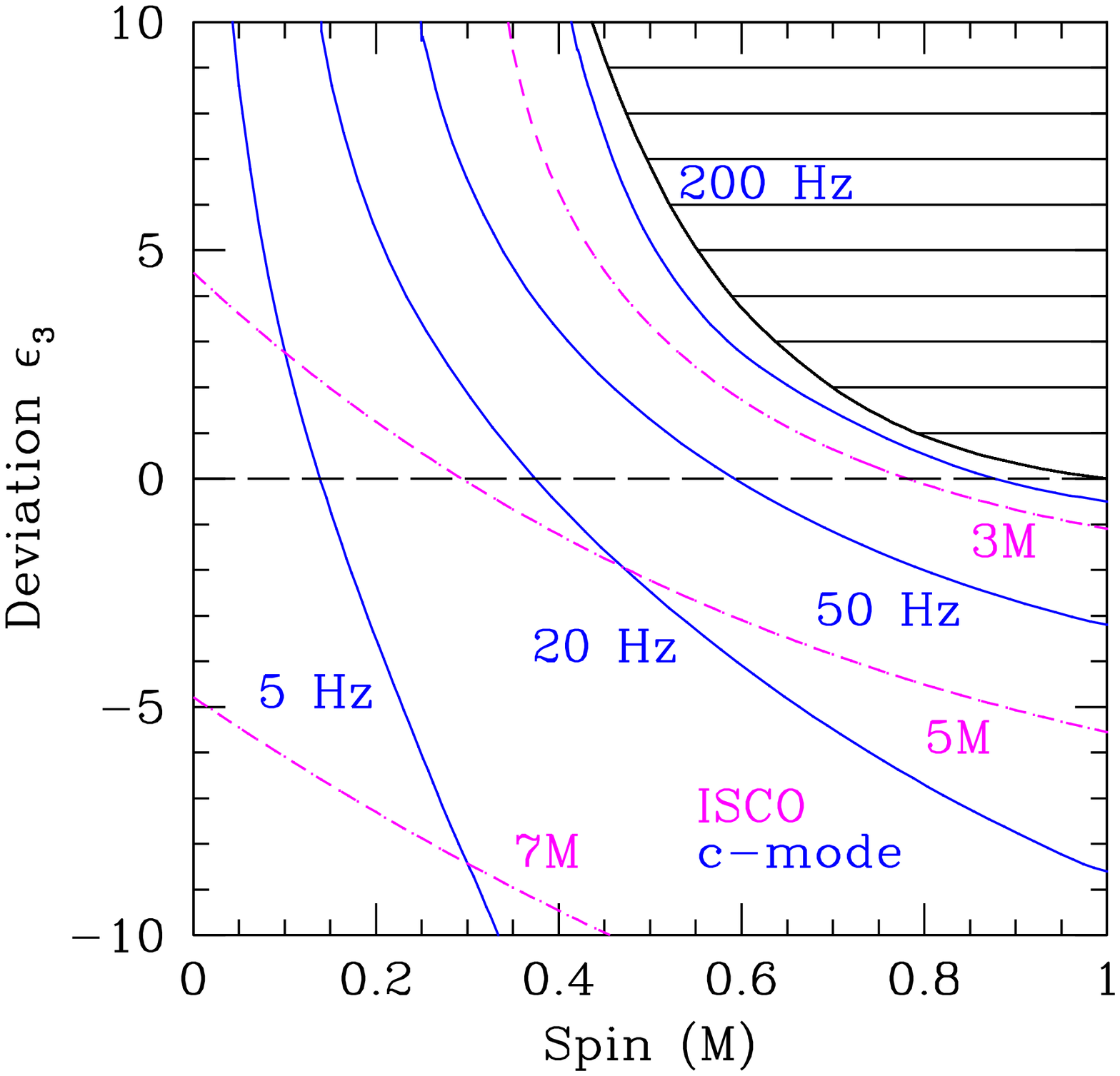,height=3.1in}
\end{center}
\caption{Left: Iron line profiles for a black hole with spin $a=0.3M$, inclination $\theta_0=30^\circ$ and emissivity index $\alpha=2$ for several values of the parameter $\epsilon_3$. Violations of the no-hair theorem manifest predominantly in the blue-shifted peak and at low energies. Right: C-mode frequency contours for a 10$M_\odot$ black hole as a function of the spin and the parameter $\epsilon_3$. These frequencies increase with increasing values of the spin and decreasing values of the parameter $\epsilon_3$. Contours of constant ISCO radius (dashed lines) are shown for comparison (Johannsen \& Psaltis 2012).}
\label{applications}
\end{figure}

Accretion flows of black holes allow us to observe potential violations of the no-hair theorem across the electromagnetic spectrum (Johannsen \& Psaltis 2010a). Observables include (a) the direct imaging of the event horizon with Sgr~A* (Johannsen \& Psaltis 2010b; Johannsen 2012; Johannsen et al. 2012b) with the {\em Event Horizon Telescope}, a planned global very-long baseline interferometer (Doeleman et al. 2009a, b; Fish et al. 2009), (b) relativistically broadened iron lines (Johannsen \& Psaltis 2012), (c) quasi-periodic variability (Johannsen \& Psaltis 2011a; Bambi 2012a), (d) accretion disk spectra (Bambi \& Barausse 2011; Krawczynski 2012), (e) X-ray polarization (Krawczynski 2012), and (f) jets (Bambi 2012b).

In Figure~2 I plot profiles of iron lines with the characteristic double-horn shape, which is produced by relativistic effects, for a black hole with a value of the spin $a=0.3M$ and several values of the parameter $\epsilon_3$. The accretion disk is assumed to be geometrically thin and optically thick with a radial extent ranging from the ISCO to $r=15M$. The emission is isotropic with an emissivity $\propto r^{-\alpha}$, $\alpha=2$, and the disk is viewed at an inclination angle of $i=30^\circ$. Violations of the no-hair theorem manifest as significant alterations of the line shape.

Figure~2 also shows frequency contours of so-called corrugation modes (c-modes; Silbergleit et al 2001) as a viable model of the quasi-periodic variability observed in both stellar-mass and supermassive black holes (e.g., Remillard \& McClintock 2006). The effects of potential violations of the no-hair theorem on iron line profiles and quasi-periodic variability may be observed with upcoming or future X-ray missions such as {\it Astro-H}, {\it ATHENA} and {\it LOFT}.

Tests of the no-hair theorem and of general relativity that are based on observations of the accretion flows of black holes require a carefully designed parametric deviation from the Kerr spacetime. The metric by Johannsen \& Psaltis (2011b) provides the necessary framework and is ideally suited for the description of rapidly-spinning black holes, which are mostly found in nature (Brenneman \& Reynolds 2009; McClintock et al. 2011). Observational tests of the no-hair theorem can be realized in the near future.

\end{document}